\documentclass[runningheads]{svmult}

\usepackage{makeidx}   
\usepackage{graphicx}  
\usepackage{subeqnar}  
\usepackage{multicol}  
\usepackage{physprbb}  
\makeindex             

\begin{document}
\title*{Angular momentum redistribution and the evolution and 
morphology of bars}
\toctitle{Evolution and morphology of bars}
\titlerunning{Evolution and morphology of bars}
%
\author{E. Athanassoula\inst{1}
}
\authorrunning{E. Athanassoula}
%
%
\institute{Observatoire de Marseille, 2 Place Le Verrier, 13248 Marseille 
cedex 04, France}

\maketitle              

\begin{abstract}
Angular momentum exchange is a driving process for the evolution of barred 
galaxies. Material at resonance in the bar region loses angular momentum 
which is taken by  material at resonance in the outer disc and/or the 
halo. By losing 
angular momentum, the bar grows stronger and slows down. This evolution 
scenario is backed by both analytical calculations and by $N$-body 
simulations. 
The morphology of the bar also depends on the amount of angular momentum 
exchanged.
\end{abstract}

\section{Introduction}
Bars are common features of disc galaxies. De Vaucouleurs~(\cite{GdV}), 
using a classification based on images at optical wavelengths, found that roughly 
one third of all disc galaxies are barred (family SB), while yet another 
third have small bars or ovals (family SAB). Observations in the near infrared 
have shown that galaxies that had been classified as non-barred from
images at optical 
wavelengths may have a clear bar component when observed in the near 
infrared. Thus Eskridge et al.~(\cite{Eskr-2000}) classified more than 70\%  
of all disc  
galaxies as barred, while Grosb{\o}l, Pompei \& Patsis~(\cite{GPP}) 
found that only $\sim5\%$ of all disc galaxies are definitely non barred. 

Bars come in a large variety of strengths, lengths, masses, axial
ratios and shapes. Great efforts have been made in order to obtain
some systematics on bar structure and important advances have been
made. Elmegreen \& Elmegreen~(\cite{E+E}) have shown that earlier type
bars are relatively longer (i.e. relative to the disc diameter) on
average than bars in later type galaxies. They also 
find that early type bars have flat intensity profiles along the bar
major axis, while late type bars have exponential-like profiles. A
correlation has been found (\cite{EA.LM}, \cite{Martin}) between the
length of bars and the size of bulges. This is in good agreement with
the trend found in \cite{E+E}, since earlier type galaxies have larger
bulges than late types. Important differences between early and late
type bars are also found with the Fourier decomposition of the surface
density. Indeed the relative $m$ = 2 and 4 components are much 
stronger in early than in late type bars. Moreover, the higher
order components ($m$ = 6 and 8), which for the late type bars are  
negligible, are still important for early types. Finally, the
shape of the bar isodensities differ and Athanassoula et
al. (\cite{AMW+}), using a sample of strongly barred early type
galaxies, showed that their bar isophotes are rectangular-like,
particularly in the region near the end of the bar. 

The first trials of $N$-body simulations~(e.g. \cite{MPQ}) show that
bars grow spontaneously and are long-lived. Yet
it is only recently that simulations have achieved sufficient quality
to provide information on the morphology of $N$-body bars and on
the mechanisms that govern bar formation and bar
evolution. I will here discuss some of the latest results of $N$-body
simulations. I will argue that it is the
exchange of angular momentum within the galaxy that will determine the
bar strength and the rate at which the pattern speed decreases after the
bar has formed, as well as the bar morphology.

\section{Angular momentum exchange and bar evolution}
Exchange of energy and angular momentum between stars at resonance with 
a spiral density wave has been first discussed by 
Lynden-Bell \& Kalnajs~(\cite{LBK}). Using linear perturbation theory, 
these authors showed that, for a steady forcing, stars emit, or 
absorb, angular momentum only if they are at resonance. Stars at the 
inner Lindblad resonance (hereafter 
ILR) lose angular momentum, while stars at the outer Lindblad resonance 
(hereafter OLR) gain it. This ground-braking work has to be extended in 
order to be applied to bars in general and $N$-body bars in particular. 
HI observations, basically starting with~\cite{Bosma}, have now 
established that, if Newton's law of 
gravity is valid, then disc galaxies are embedded in a dark 
matter component, called the halo, whose mass exceeds that of the disc. 
This component should now be taken into account as an extra partner in 
the angular momentum exchange process. Furthermore, bars are strongly 
non-linear 
features, since they contain a large fraction of the mass in the inner 
parts of the disc and a considerable part of the total disc mass. Thus 
any linear theory should be thought of as a guiding line, to be supplemented 
by and tested against adequate $N$-body simulations. It is obvious that 
such simulations should be fully self-consistent, since rigid components 
can not exchange energy and angular momentum. 

In~\cite{EA.Jmom} I extended the analytical work of~\cite{LBK} to include 
spheroidal components, like a halo and/or a bulge, and also supplemented it 
with fully self-consistent $N$-body simulations. In the analytical part 
I showed that, if the distribution function of the spheroidal
component is a function only of the energy, then at all resonances the
halo and bulge particles can only gain  
angular momentum. Also, since the bar is a strongly nonlinear feature, higher 
multiplicity resonances should be taken into account. Thus angular 
momentum is emitted by particles (stars) at the resonances in the inner 
disc, mainly the ILR, but also the inner -1:m resonances nearer to corotation 
(hereafter CR). It is absorbed by disc particles (stars) at the OLR, or the 
outer 1:m resonances, outside corotation, or by the resonant particles in 
the halo and/or bulge components. Since the bar is a negative angular 
momentum perturbation~(\cite{LBK}), by losing angular momentum it will grow.
This clearly outlines a scenario for the evolution of barred galaxies.

\begin{figure}[ht]

\vspace{-0.0cm}

\begin{center}
\includegraphics[width=.5\textwidth]{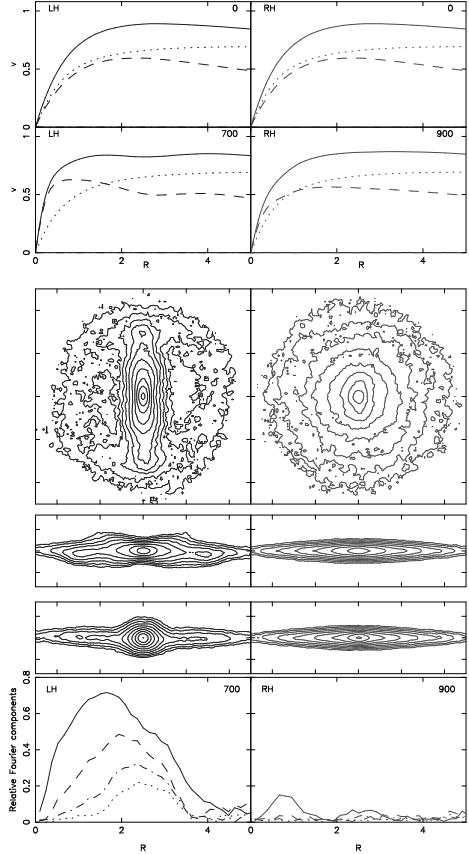}
\end{center}
\caption[]{Basic information on simulations LH (time $t$ = 600) and RH 
(time $t$ = 900).
The two upper rows give the circular
velocity curves at time 0 and $t$. The dashed and dotted
lines give the contributions of the disc and halo
respectively, while the thick full lines give the total circular
velocity curves. The third row of panels gives the isocontours of the
density of the disc particles projected face-on and the fourth and
fifth row give the side-on and end-on edge-on views, respectively.
The side of the box for the face-on views is 10 length units and the height
of the box for the edge-on views is 3.33. The isodensities in the third row of
panels have been chosen so as to show best the
features in the bar and in the inner disc. No isodensities for the
outer disc have been included, although the disc extends beyond the
area shown in the figure. The sixth row of panels gives the
$m$ = 2, 4, 6, and 8 Fourier components of the mass.
}
\label{fig:basic_halo}
\end{figure}

\section{The effect of the halo on bar evolution}

As the bar loses angular momentum, it grows stronger. This, however,
can only happen if there are absorbers that can absorb the  
angular momentum that the bar region emits. Thus the existence of a
massive halo component, whose  
resonances can absorb considerable amounts of angular momentum, will help 
the bar grow. At first sight this may be thought to go against old 
claims that haloes stabilise bars. In fact, a more precise wording is
necessary. Indeed, the halo slows down 
the bar growth in the initial stages of the evolution. At later stages, 
however, the situation can be reversed, since the halo may absorb the
angular momentum emitted by the bar, and thus it may allow the latter 
to grow further. Thus bars that grow in halo-dominated discs can be 
stronger than bars that grow in disc-dominated surroundings. This 
effect was not seen till recently, since the older studies  
were either 2D (e.g. \cite{Toomre},~\cite{EA.JAS}), or 3D but with few 
particles (e.g. \cite{Ostr.Peeb}), or had rigid haloes.  
In all these cases 
the halo was denied from the onset its destabilising influence. Its effect 
becomes clear in fully self-consistent $N$-body simulations, with an adequate 
particle number and resolution. Thus~\cite{EA.AM} showed that stronger bars 
can grow in cases with more important halo components. 

The influence of the halo is also 
illustrated in Fig.~\ref{fig:basic_halo}, where I compare the results 
of two $N$-body simulations. Initially their disc is exponential, with
unit mass and scale-length ($M_d$ = 1, $R_d$ = 1) and its 
$Q$ parameter (\cite{ToomQ}) is independent of radius and equal to
1.2. Since $G$ = 1, taking the mass of the disc equal to 5 $\times$
$10^{10}$ M$_\odot$, and its scale-length equal to 3.5 kpc implies that the
unit of velocity is 248 km/sec and the unit of time is 1.4 $\times$
$10^7$ yrs. This calibration is reasonable, but is not unique, so in
the following I will give all quantities in computer units. The reader
can then convert the values to astronomical units according to his/her
needs. The halo component is spherical, non-rotating and has an
isotropic velocity distribution. It follows a pseudo-isothermal 
radial density profile (\cite{hernq}) and has a total 
mass $M_h$ = 5, a core radius $\gamma$ = 0.5 and a cutoff radius
of $r_c$ = 10. Its density is truncated at 15 disc scale-lengths, 
i.e. at a radius containing more than 96\% of its mass. In building the initial
conditions I loosely followed~\cite{hernq} and \cite{EA.AM}, and the 
simulations were
run on the Marseille GRAPE-5 systems (for a description of the GRAPE-5
boards see~\cite{Kawai}). The only difference between the initial
conditions of the two
simulations is that for simulation LH, illustrated in the left panels, 
the halo is live and represented by roughly $10^6$ particles, while 
for simulation RH, illustrated in
the right panels, it is rigid, i.e. represented by an analytical
potential and thus can 
neither emit nor absorb angular momentum. Although their initial 
conditions are so similar, the two simulations evolve in a very
different way. After some initial multi-spiral episodes, LH forms a
bar which grows stronger with time. Its morphology at $t$ = 700 is shown 
in the left panels. The bar is long and strong and has ansae-type
features near the end of its major axis. It is surrounded by a ring,
which can be compared to the inner rings often observed in barred
galaxies. The bar formation entails considerable redistribution of the disc
matter, both radially and azimuthally. On the other hand the disc in
simulation RH stays close to 
axisymmetric, except for some multi-armed spirals which die out with
time. Only at the latest stages of the evolution does it form an oval
distortion, and even that is weak and short and is confined to the
innermost parts of the disc, as can be seen for $t$ = 900 in the right
panels of Fig.~\ref{fig:basic_halo}. I show this simulation at a 
later time than that for simulation LH because at earlier times 
there is very little structure visible.    
  
Seen edge-on with the bar seen side-on (i.e. with the line of sight
along the bar minor axis), simulation LH exhibits a very strong peanut, 
which is totally absent from simulation RH 
(fourth row of panels). Seen edge-on with the bar seen end-on
(i.e. with the line of sight along the bar major axis), the peanut in
LH resembles a bulge (left panel on fifth row). This underlines the
hazards involved in classifying edge-on galaxies, since the classifier
may in such cases easily misinterpret the bar for a bulge.

A useful way of quantifying the bar strength is with the help of the
Fourier components of the mass, or density. These can be defined as

\begin{equation}
F_m (r) = \sqrt{A_m^2(r)+B_m^2(r)}/ A_0(r), ~~~~~~~~~~m=0, 1, 2, .....
\end{equation}

\noindent
where

\begin{equation}
A_m (r) = \frac {1}{\pi} \int _0 ^{2 \pi} \Sigma (r, \theta) cos (m \theta) d \theta, ~~~~~~~~~m=0, 1, 2, .....
\end{equation}

\noindent
and

\begin{equation}
B_m (r) = \frac {1}{\pi} \int _0 ^{2 \pi} \Sigma (r, \theta) sin (m \theta) d \theta, ~~~~~~~m=1, 2, .....
\end{equation}

\noindent
For runs LH and RH, these components for $m$ = 2, 4, 6 and 8 are shown
in the lower panels of Fig.~\ref{fig:basic_halo}. For run LH all four
components are important, due to the strength of the bar. Their
amplitude decreases with increasing $m$, while the location of the
maximum shifts outwards. On the other hand, for model RH only the $m$ =
2 component stands out from the noise, but its amplitude is rather
small, smaller than e.g. that of the $m$ = 8 for model LH.

Since the only difference between the initial conditions of models LH
and RH is that the halo of the one is responsive, while that of the
other is rigid, we can conclude that the halo response is crucial for
determining the evolution of barred galaxies.

\section{Bar slow-down}

\begin{figure}[ht]
\begin{center}
\includegraphics[width=.6\textwidth]{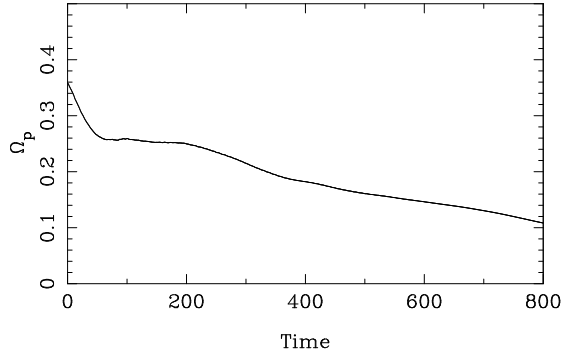}
\end{center}
\caption[]{Bar pattern speed of simulation LH as a function of time. 
}
\label{fig:omegap_halo}
\end{figure}

I ran a large number of simulations similar to those described in the 
previous sections. I noted that, as it loses angular momentum, the 
bar grows longer, 
and more massive, thus stronger. Angular momentum loss, however,  
is not only linked to an increase in the bar strength. It is also linked
to a slow-down, i.e. to a decrease of the bar pattern speed $\Omega_p$ with 
time. Such a slow-down has indeed been seen in a number of simulations 
and has also been predicted analytically (\cite{TW}, \cite{W}, 
\cite{LC1}, \cite{LC2}, \cite{HW}, \cite{EA.Buta}, \cite{DS1},
\cite{DS2}). It can also be  
seen in Fig.~\ref{fig:omegap_halo}, which shows the run of the 
bar pattern speed with time for simulation LH, whose morphology at
time $t$ = 700 is 
shown in the left panels of Fig.~\ref{fig:basic_halo}. Note 
that the bar slows down considerably with time.

\section{Resonances}

\begin{figure}[ht]
\begin{center}
\includegraphics[width=.85\textwidth]{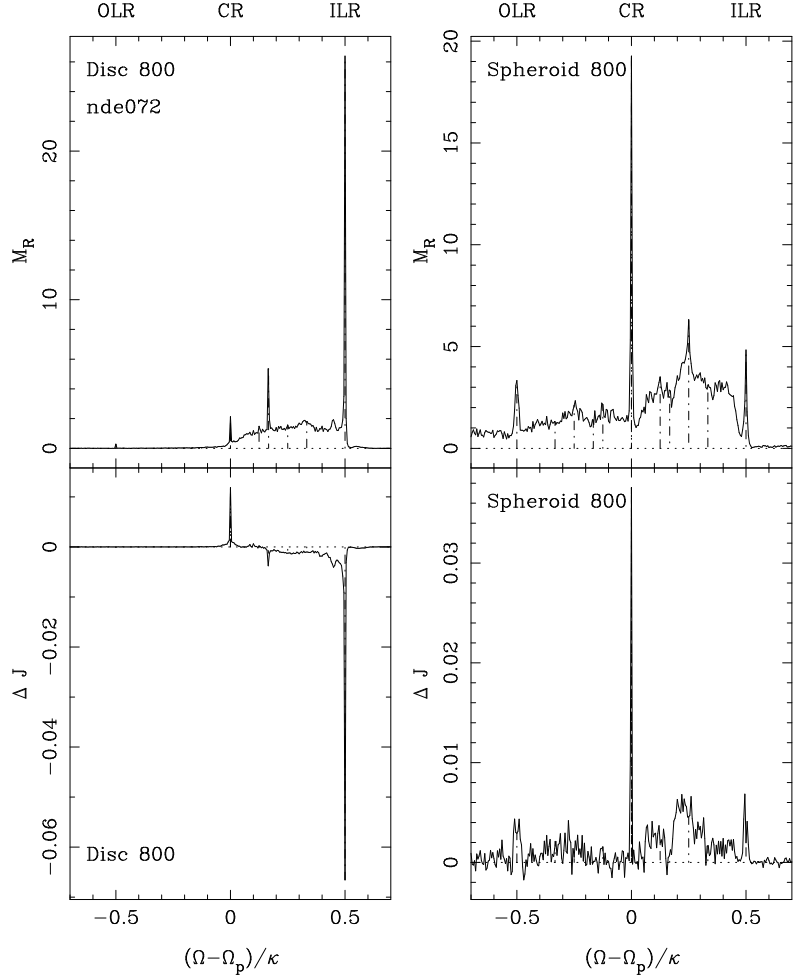}
\end{center}
\caption[]{The upper panels give, for time $t$ = 800, the mass per 
unit frequency ratio, $M_R$, as a
function of that ratio. The lower panels give $\Delta J$, the angular momentum
gained or lost by the particles between times 800 and 500, plotted as
a function of their frequency ratio $(\Omega - \Omega_p) / \kappa$,
calculated at time $t$ = 800. The left panels correspond to the disc
component and the right ones to the halo. The
component and the time are written in the upper left corner of each
panel. The vertical dot-dashed lines give the positions of the main
resonances.}
\label{fig:resonances}
\end{figure}

In order for haloes to be able to absorb angular momentum, they need 
to have a considerable fraction of their mass at resonance. This
was shown to be true in (\cite{EA.ApJL}). I will illustrate it here 
for model LH.
The procedure is the same as that followed in~\cite{EA.ApJL}. 
I calculate the potential from the mass distribution 
in the disc and halo component at time $t$ = 800, by freezing all
motion except for the bar, to which I assign bulk rotation with
a pattern speed equal to that found in the simulation at that time. 
I then pick at random 100\,000 disc and 100\,000 halo particles and, 
using their positions and velocities as initial conditions, I calculate 
their orbits for 40 bar rotations. Using spectral 
analysis (\cite{BS}, \cite{Laskar}), I then find the principal 
frequencies of these orbits, i.e. the angular velocity 
$\Omega$, the epicyclic frequency $\kappa$ and the vertical frequency 
$\kappa_z$. An orbit is resonant if there are three integers $l$, 
$m$ and $n$, such that

\begin{equation}
l\kappa+m\Omega+n\kappa_z = - \omega_R = m \Omega_p 
\end{equation}

\noindent
Orbits on planar resonances fulfill

\begin{equation}
l\kappa+m\Omega = - \omega_R = m \Omega_p
\label{eq:resondef} 
\end{equation}

\noindent
The ILR corresponds to $l$ = --1 and $m$ = 2, CR to  $l$ = 0, and OLR
to $l$ = 1 and $m$ = 2. 

The upper panels of Fig.~\ref{fig:resonances} show, for time $t$ =
800, the mass per unit frequency ratio $M_R$ of particles having a given 
value of the 
frequency ratio $(\Omega - \Omega_p) / \kappa$ as a function of this
frequency ratio\footnote{See \cite{EA.Jmom} for more information on 
$M_R$ and on how it is derived.}. The distribution is not uniform, 
but has clear peaks
at the location of the main resonances, as was already shown in
~\cite{EA.ApJL} and ~\cite{EA.Jmom}. The
highest peak for the disc component is at the ILR, followed by
$(-1, 6)$ and CR. In all simulations with strong bars 
the ILR peak is strong. The existence of peaks at other resonances as
well as their importance varies from one run to
another and also during the evolution of a given run. For example the
CR peak is, in many other simulations, much stronger than in the
example shown here. For the
spheroidal component the highest peak is at CR, followed by peaks 
at the ILR, OLR and $(-1, 4)$. 

The lower panels show the angular
momentum exchanged. For this I calculated the angular momentum of each
particle at time 800 and at time 500, as described in
~\cite{EA.Jmom}, and plotted the difference as a
function of the frequency ratio of the particle at time
800. It is clear from the figure that disc particles at ILR and at the
$(-1, 6)$ resonance lose angular momentum, while those at CR gain it. 
There is a also a general, albeit small, loss of angular 
momentum from particles with frequencies between CR and ILR. This
could be partly due to particles 
trapped around secondary resonances, and partly due to angular
momentum taken from particles which are neither resonant, nor
near-resonant, but can still lose a small amount of angular momentum
because the bar is growing. The corresponding panel for
the spheroid is, as expected, more noisy, but shows that particles at all
resonances gain angular momentum. Thus this plot, and similar ones
which I did for other simulations, confirm the analytical results
of~\cite{EA.Jmom}, and show that the linear results concerning the 
angular momentum gain or loss by resonant particles, 
qualitatively at least, carry over to the strongly nonlinear regime.

\section{What determines the strength of bars and their slow-down rate?}
I have shown in the previous sections that the halo can take angular 
momentum from the bar, thus making it stronger and slower. However, 
for this effect to be important, the amount of angular momentum exchanged 
must be considerable. For the latter to happen the halo must

\begin{itemize}
\item
be sufficiently massive in the regions containing the principal resonances.

\item
not be too hot, i.e. not have too high velocity dispersion. Indeed, hot
haloes can not absorb much angular momentum, even if they are massive 
(e.g.~\cite{EA.Jmom}).

\end{itemize}

Thus the length and the slow-down rate of bars are naturally limited 
by the mass and velocity distribution of the halo. Examples of this 
can be found in~\cite{EA.Jmom}.
  
\section{Trends and correlations}
\label{sec:correl}

\begin{figure}[ht]
\begin{center}
\includegraphics[width=.8\textwidth]{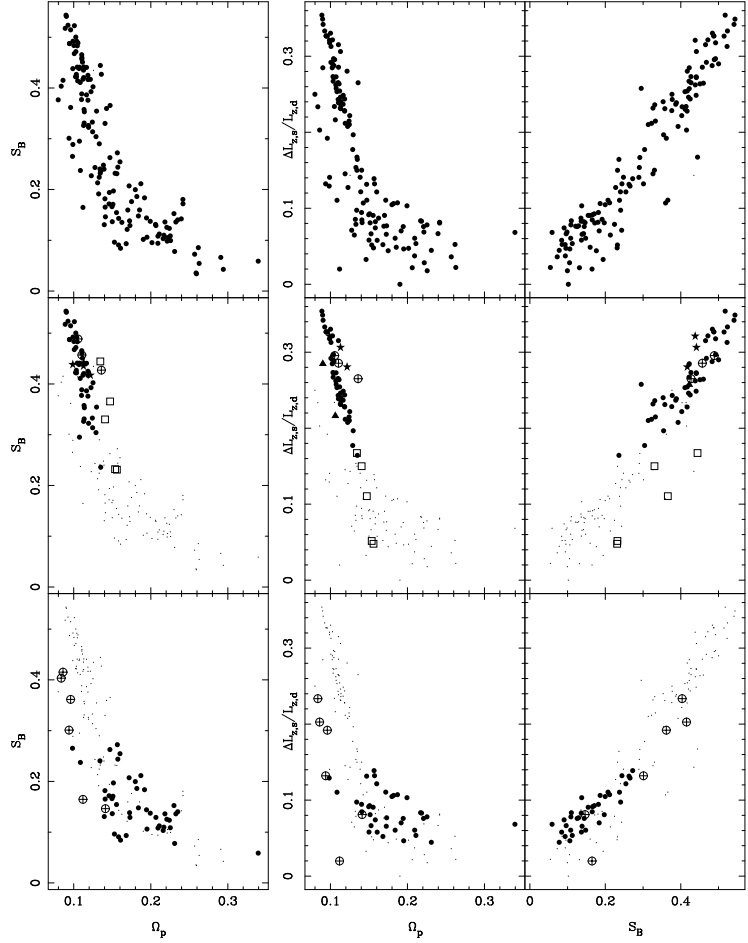}
\end{center}
\caption[]{Relations between the bar strength and the pattern speed
(left panels), the spheroid angular momentum and the pattern speed (middle
panels) and the spheroid angular momentum and the bar strength (right
panels), at times $t = $ 800. The spheroid angular momentum is
normalised by the initial disc angular momentum ($L_{z,d}$). The
simulations under consideration in each panel are
marked with a filled circle and the rest by a dot. The upper row
includes all simulations, the middle and the lower ones
subsamples, as described in the text. In the
  middle panel simulations with a bulge are marked with a $\oplus$,
simulations with $\gamma =$ 0.01 with a filled star, simulations with  
$2 > \gamma \ge$ 1 with a filled triangle and simulations with $Q_{init}
\ge$ 2 with an open square. In the lower panel simulations with
$Q_{init} \ge$ 1.4 and $z_0 \ge$ 0.2 are marked with an $\oplus$. 
}
\label{fig:correl}
\end{figure}

In~\cite{EA.Jmom} I found trends and correlations between the angular
momentum absorbed by the spheroid (i.e. the halo plus, whenever 
existent, the bulge), the bar strength and the bar pattern 
speed. They are based on a set of simulations analogous to those
described in the previous sections. Such plots are given also
in Fig.~\ref{fig:correl}, based on a somewhat 
larger sample of simulations. About three quarters of them were 
run with the Marseille GRAPE-5 systems, and roughly one quarter 
was run on PCs using Dehnen's treecode (\cite{D1}, \cite{D2}).
Each point represents one simulation 
and the trends are the same as those found in~\cite{EA.Jmom}.
The upper panels show the results for the whole sample, the 
middle panels contain only simulations where the halo has a small
core radius ($\gamma < 2$), $M_h$ = 5 and does not extended beyond 15 disc
scale-lengths, and the lower ones contain only  
simulations where the halo has a large core radius ($\gamma > 2$), 
$M_h$ = 5  and again does  not extended beyond 15 disc scale-lengths. 

The right panel shows that there is a correlation between the angular
momentum of the spheroid and the bar strength. This correlation holds
also when I restrict myself to simulations with large (or small) core
radii as seen in the second and third row. A trend also exists between
the spheroid 
angular momentum and the bar pattern speed. In this case, however,
simulations with large core radii behave differently from those with
small radii. Indeed, for simulations with a small core radius
(i.e. centrally concentrated halos) I find a very strong correlation
between the spheroid angular momentum and the pattern speed,
particularly if I restrict myself to one value of $\gamma$. In
such simulations the angular momentum is exchanged primarily between the 
bar region and the spheroid, thus accounting for the very tight 
correlation. Simulations with large cores behave differently (lower 
middle panel). They show only a rough trend, except for simulations 
with a hot disc, which show a tight correlation. This is easily explained 
in the scenario of evolution via angular momentum exchange. The outer 
parts of hot discs absorb only little angular momentum, so that the 
exchange is basically between the bar region and the spheroid, thus 
accounting for the tight correlation. On the other hand, if the outer 
disc is cold, then it can participate more actively in the exchange. Since the 
angular momentum absorbed by the spheroid (plotted in Fig.~\ref{fig:correl}) 
is not the total angular momentum exchanged, but only a fraction of it, 
I find only a trend.
   
\section{Comparing the morphology of $N$-body and of real bars}
\begin{figure}[t]
\begin{center}
\includegraphics[width=.97\textwidth]{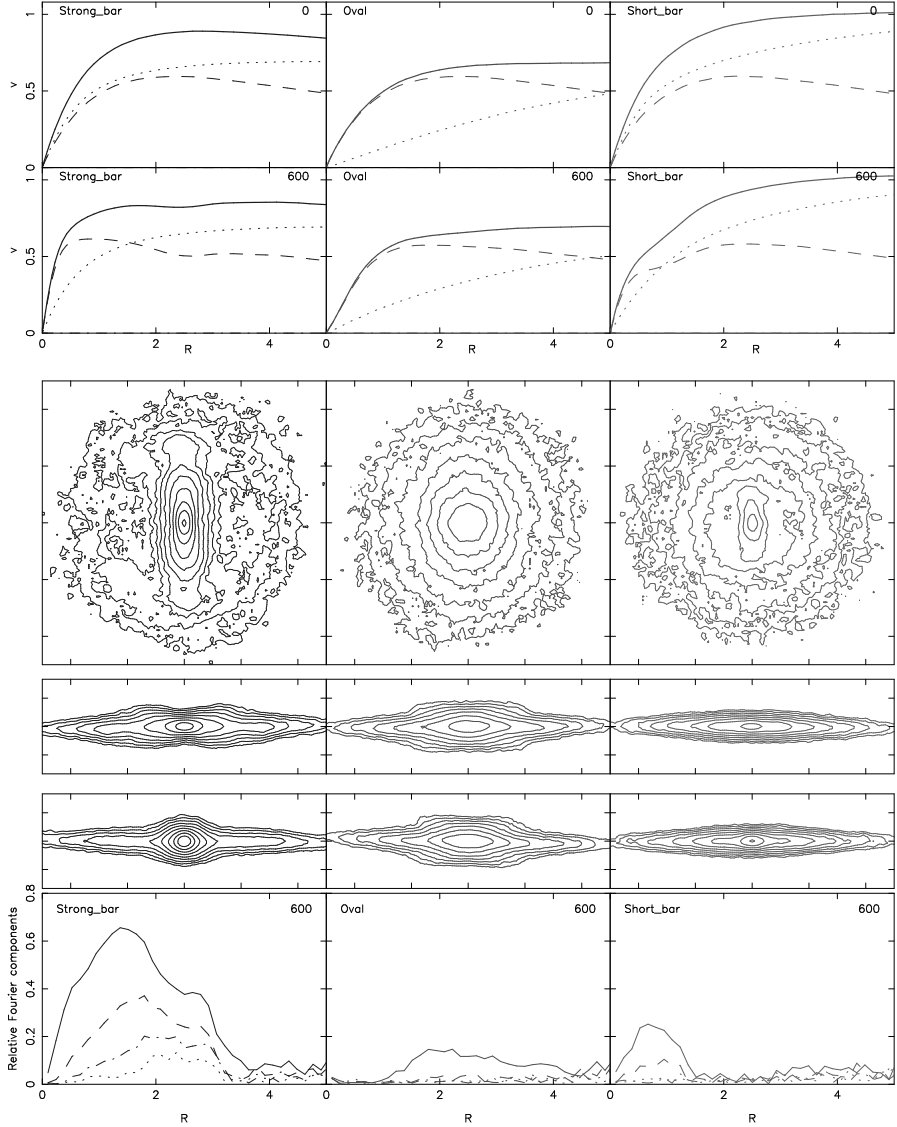}
\end{center}
\caption[]{
Comparison of a simulation forming a strong bar (left panels), one
forming an oval (middle panels) and one forming a short bar (right
panels). The layout is as for Fig.~\ref{fig:basic_halo}.
}
\label{fig:compare_basic}
\end{figure}

\begin{figure}[t]
\begin{center}
\includegraphics[width=.97\textwidth]{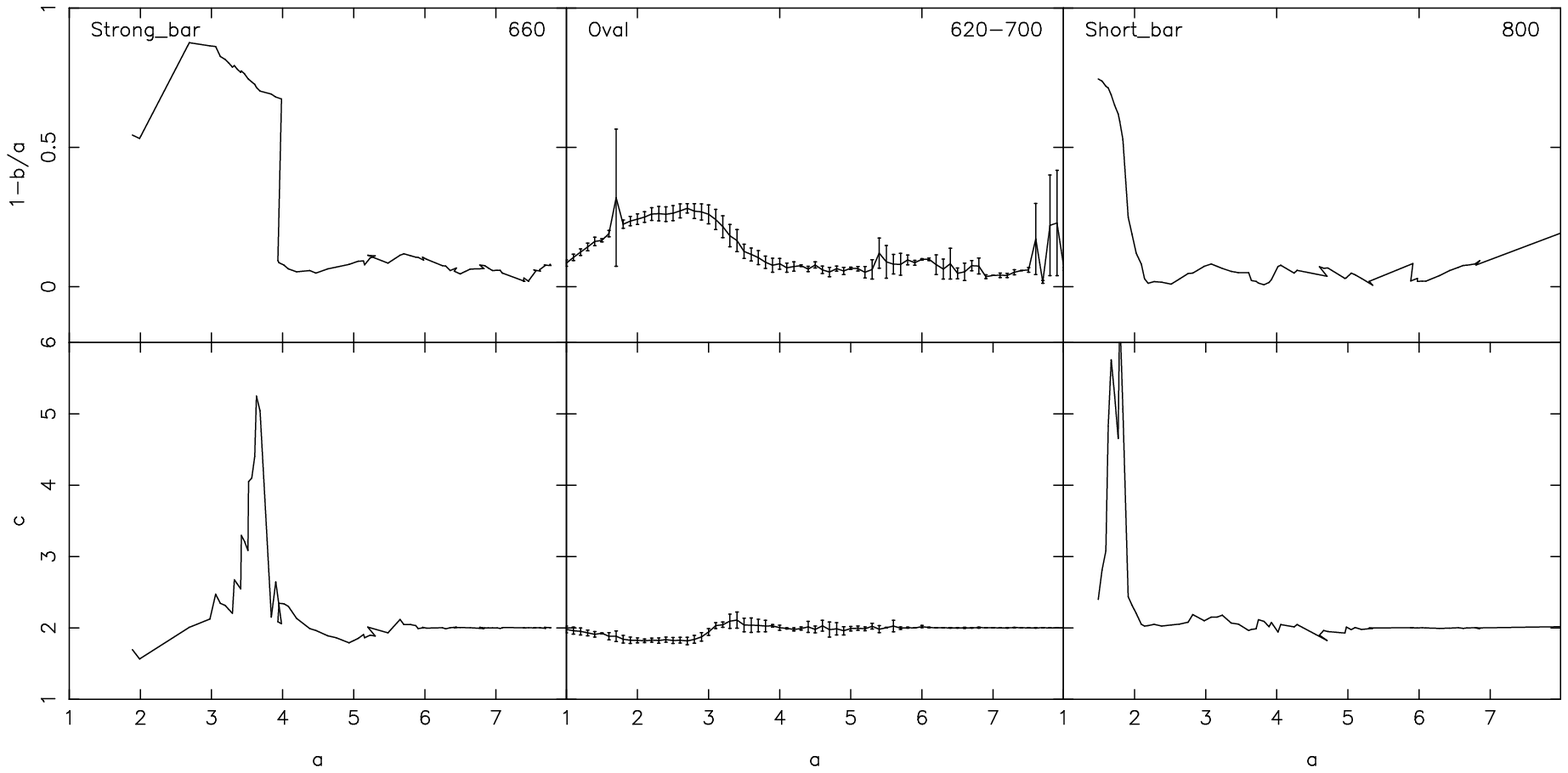}
\end{center}
\caption[]{
The upper panels show the run of the ellipticity 1 - $b/a$ as
a function of the semi-major axis $a$. The lower panels show the run of
the shape parameter $c$, also as a function of $a$. The left panels
corresponds to a model with a strong bar, the middle ones to model
with an oval and the right ones to a model with a short bar. To
improve the signal-to-noise ratio for the model with the oval I took
an average over a time interval, namely [620-700]. The dispersion during 
that time is indicated by the error bars. The times are given
in the upper right corner of the upper panels.}
\label{fig:axrat}
\end{figure}

The correlations discussed in section \ref{sec:correl} show clearly that
models that have exchanged more angular momentum have stronger bars
than models that have exchanged little. By examining the results of
the individual simulations, I could see that, in cases where large amounts
of angular momentum have been exchanged, the bars are long, relatively
thin and have rectangular-like isodensities, particularly in their
outer parts. A typical example of such a case is given in the
left panels of Fig.~\ref{fig:compare_basic} (see also
\cite{EA.AM}). Note also the existence 
of ansae at the ends of the bar, a feature sometimes observed in early
type barred galaxies.
On the other hand, models that have exchanged little angular momentum have
less homogeneous properties. For example they can have either ovals,
or short bars. Typical examples of such cases are given in the middle and right
panels of Fig.~\ref{fig:compare_basic}, respectively. The model in the
left panel has exchanged about 15 percent of the disc angular momentum
by the time shown in Fig.~\ref{fig:compare_basic}, while the other two
models only of the order of a percent.

The edge-on morphology also is strongly influenced by the amount of
angular momentum exchanged. The strong bar, when seen edge-on, 
displays a clear peanut morphology, as often observed. On the other 
hand the oval has a boxy edge-on appearance, while the small bar has 
not changed significantly the edge-on morphology of the galaxy. 

The difference in bar strength is also illustrated in the lower panels of 
Fig.~\ref{fig:compare_basic}, which show the relative Fourier components 
of the density for $m$ = 2, 4, 6 and 8 for the three simulations. The 
simulation that exchanged a lot of angular momentum has a very strong
$m$ = 2 component, with a secondary maximum roughly at the position of 
the ansae. The remaining components, even the $m$ = 6 and 8 ones, are 
also important. The location of their maximum moves outwards with 
increasing $m$. The oval has much lower Fourier components, and only 
the $m$ = 2 stands out from the noise. For small radii all components 
are nearly zero, which means that the oval must be nearly axisymmetric
in its innermost parts. On the other hand, the $m$ = 2 amplitude drops 
slowly with radius in the outer parts, thus extending to large
radii. The small bar has Fourier components which drop rapidly with
radius, i.e. they are noticeable only in the central region, as expected 
since the bar is confined there. Only
the $m$ = 2 and 4 components stand out from the noise.

The radial rearrangement of the disc material due to the bar can be 
inferred by comparing the initial with the current circular velocity
curves, given in the first and second rows of panels. The strong bar
has entailed a substantial radial rearrangement, the final disc mass
distribution being considerably more centrally concentrated than the 
initial one. On the other hand, in the other two simulations, and
particularly in the one producing the oval, there is very little
radial rearrangement of the disc material. Since there is also hardly
any radial rearangement of the halo material (\cite{EA.AM},
\cite{EA.Puebla}, \cite{OV.AK}), this means that the
disc-to-halo mass ratio changes most in the simulations where more
angular momentum has been exchanged. 

Quantitative comparison of the bar form of the three models is given 
in Fig.~\ref{fig:axrat}. The values of the bar semi-major and semi-minor 
axes ($a$ and $b$, respectively) and of the shape parameter ($c$) were 
obtained by fitting generalised ellipses of the form

\begin{equation}
(|x|/a)^c + (|y|/b)^c = 1,
\end{equation}

\noindent
to the bar isodensities. The shape parameter $c$ is 2 for ellipses, 
larger than 2 for rectangular-like generalised ellipses, and smaller 
than 2 for diamond-like ones. From this figure one can note that both the
strong and the short bar are thin, and in general see how their axial 
ratios vary with the semi-major axis. The shape parameter is given 
in the lower panels. We see that both the strong and the short bar have 
rectangular-like isodensities in the outer regions of the bar, while 
the oval has a shape very close to elliptical. In fact the strong bar
has axial ratios and shapes very similar to those found in \cite{AMW+} 
by applying the same type of analysis to a sample of early type barred 
galaxies.  

Plotting the run of the density along the bar major axis (\cite{EA.AM})
I find for the strong bar a profile which is rather flat within the 
bar region, similar to what was found in \cite{E+E} for early type bars.

It is thus clear that the amount of angular momentum exchanged influences 
the morphology of the bar. In my first example, where a lot of angular
momentum was transferred from the bar to the outer halo (mainly), the 
result is a long, strong bar, with some rectangular-like isophotes and 
ansae at its ends. The examples where little angular momentum was
exchanged have a very different morphology, one forming an oval and 
the other a short bar. What determines which one of the two it will 
be? In the examples shown here, and in a rather large sample of 
similar cases, the oval was formed in an initially hot disc, while 
the short bar grew in a hot halo. The existing theoretical framework, 
however, gives no predictions on this point and work is in progress
to elucidate this further.  
 
\section{Summary}
In this paper I reviewed evidence that shows that angular momentum 
is exchanged between the bar region in the one hand, and the outer
disc and the spheroid on the other. This exchange determines the 
slow-down rate of the bar, as well as its strength and its overall
morphology.

\medskip
\noindent
{\bf Acknowledgments.} I thank the organisers for inviting me to
this interesting conference. I thank M. Tagger, A. Bosma,
W. Dehnen, A. Misiriotis, C. Heller, I. Shlosman, F. Masset,
J. Sellwood, O. Valenzuela, A. Klypin and P. Teuben for many
stimulating discussions. I thank J.~C. Lambert for his help with the
GRAPE 
software and with the administration of the simulations and W. Dehnen for
making available to me his tree code and related programs. I
also thank the INSU/CNRS, the University of Aix-Marseille I, the Region PACA
and the IGRAP for funds to develop
the GRAPE and Beowulf computing facilities used for the simulations
discussed in this paper and for their analysis. 
\vskip 0.5cm
%

%

\end{document}